# Direct Patterning of Colloidal Quantum-Dot Thin Films for Enhanced and Spectrally Selective Out-Coupling of Emission

Ferry Prins[*], David K. Kim[*,†], Jian Cui, Eva De Leo, Kevin M. McPeak[§], and David J. Norris

Optical Materials Engineering Laboratory, Department of Mechanical and Process Engineering,

ETH Zurich, 8092 Zurich, Switzerland

We report on a template-stripping method for the direct surface patterning of colloidal quantum-dot thin films to produce highly luminescent structures with feature sizes less than 100 nm. Through the careful design of high quality bull's-eye gratings we can produce strong directional beaming (10° divergence) with up to six-fold out-coupling enhancement of spontaneous emission in the surface-normal direction. A transition to narrow single-mode lasing is observed in these same structures at thresholds as low as 120 µJ/cm$^2$. Furthermore, making use of the size-tunable character of colloidal quantum dots, we demonstrate spectrally selective out-coupling of light from mixed quantum-dot films. Our results provide a straightforward route towards significantly improved optical properties of colloidal quantum-dot assemblies.

[*] these authors contributed equally
To whom correspondence should be addressed: dnorris@ethz.ch (D.J.N.)
[†] Current address: MIT Lincoln Laboratory, Lexington, MA, USA
[§] Current address: Cain Department of Chemical Engineering, Louisiana State University, Baton Rouge, LA, USA

Colloidal quantum dots (cQDs), or semiconductor nanocrystals, are highly versatile building blocks that combine size-tunable optical properties with low-cost wet-chemical processing.[1,2] High quantum yields (>90%)[3] and spectrally-narrow emission throughout the visible-near-infrared range[4] have placed cQDs among the highest color-quality emitters available, while at the same time outperforming most organic emitters in terms of photo-stability.[5] A number of cQD-based display[6,7] and lighting technologies[8–10] have recently emerged, while backlit display systems have already entered the market.[11] With internal conversion efficiencies of cQDs approaching unity, the light-extraction efficiencies from cQD films has become the limiting factor in terms of brightness. As is inherent to such high refractive-index layers, a significant portion of the emission undergoes internal reflection and gets trapped into waveguiding modes. To overcome this issue, research efforts are increasingly shifting toward the development of nanophotonic light-management strategies that improve the out-coupling from these materials.

A number of successful photonic strategies for enhanced extraction of cQD emission have been developed recently. These include, for example, external out-coupling schemes in which the cQDs are coupled to plasmonic[12] or photonic-crystal[13] structures, or are embedded in photonic microcavities.[14] While these external out-coupling schemes can provide strongly enhanced[15] and highly directional[12] extraction of cQD emission, it remains challenging to achieve these characteristics without significantly increasing the complexity of the overall system. A conceptually more straightforward approach would be to pattern the high refractive-index cQD film directly. Direct patterning of the light-emitting layer has been successfully applied in organic light-emitting polymers, for example by structuring the surface with linear or circular Bragg gratings.[16,17]

Bragg scattering of waveguided light leads to constructive interference of emission in the out-of-plane direction when the in-plane momentum is conserved. The angle of out-coupling $\theta$ with respect to the surface normal for a given emission wavelength $\lambda_{em}$ and a grating periodicity $\Lambda$ is then governed by the Bragg condition, given by

$$k_0 \sin \theta = \pm k_{wg} \pm m\, k_g = \pm \frac{2\pi n_{eff}}{\lambda_{em}} \pm m \frac{2\pi}{\Lambda}, \qquad (1)$$

where $k_0$ is the wavevector of the outcoupled light, $k_{wg}$ is the wavevector of the waveguided light, $k_g$ is the Bragg vector, $n_{eff}$ is the effective refractive index of the waveguide mode, and $m$ is the diffraction order. Hence, first order diffraction allows waveguided light to scatter in the out-of-plane ($\theta = 0$) normal direction when $\lambda_{em} = \Lambda \cdot n_{eff}$. In a circularly symmetric bull's-eye grating, where momentum matching is satisfied for each radial direction, this results in a distinct beaming of fluorescence from the center of the structure.[18] Additionally, while the first-order diffraction couples the emission out vertically, the second-order diffraction can provide feedback for the in-plane fields - a concept that has been successfully employed to construct surface-emitting Bragg lasers.[19]

The fabrication of Bragg gratings requires high-quality sub-wavelength patterning of the surface of the light-emitting film. In organic films, Bragg gratings have been fabricated using for example soft lithography or stamping.[16,19,20] Even though cQDs profit from many of the advantages of solution processability, few methods exist to produce patterned assemblies with submicron resolution.[21] Inkjet[22] and electrohydrodynamic[23] printing techniques have been developed to produce patterned films, and are capable of achieving submicron resolution. However, the structural definition of taller structures is poor for these techniques, making them unsuitable for high-quality photonic Bragg gratings. Soft lithography and stamping techniques do exist for cQD solids,[7,24] but submicron resolution is challenging. In particular, the reduced structural cohesion of cQD solids as compared to most organic materials makes patterning of cQD solids using soft lithography more prone to pattern distortion. Pattern definition using electron beam lithography in combination with lift-off techniques would potentially allow for higher-resolution patterning of taller structures.[25] However, while such techniques have yielded patterned cQD films for electronic applications,[25] photonic structuring has not yet been reported. Finally, alternative strategies for printing,[26] patterning,[27] and soft-lithography[28,29] have been developed using matrix-embedded cQD films, though the use of such composite films has the distinct disadvantage of diluted emitter concentration, which inevitably places limitations on the emission intensity.

Here, we present a new methodology for the direct patterning of cQD assemblies using template stripping, yielding significant improvements in the out-coupling efficiency of cQD emission using surface-patterned bull's eye gratings. Our fabrication method is similar to template stripping techniques used to fabricate patterned metal films[30] and macroscale assemblies of plasmonic nanoparticles.[31] Figure 1a shows a schematic of the process. Patterned silicon templates are fabricated using standard electron-beam lithography and reactive ion etching, and are subsequently coated with a self-assembled monolayer using a previously reported[32,33] octadecyltrichlorosilane (ODTS) treatment (see Methods for details). The cQDs are then drop-cast onto the template to form a dense and continuous layer. In the next step, a glass backing substrate is attached to the colloidal film using an ultraviolet-curable epoxy. In a final step, the colloidal film is mechanically peeled off the template to expose the patterned surface defined by the template. The template can be re-used after a simple cleaning procedure. The use of the high-quality self-assembled monolayer on the silicon template is essential to the success of this technique. Template stripping relies on a well-defined "weakest link" in the layer stack. Without the self-assembled monolayer, the adhesion between the cQDs and the template would be competing with the cQD-cQD adhesion in the film itself.[7] Using our technique, lateral feature sizes as small as 100 nm are achievable, while feature heights as tall as 100 nm were reproducibly obtained (see Supporting Information).

Making use of the size-tunable character of the quantum dots, films of different colors can be prepared. Figure 1b shows a photograph of ultraviolet-illuminated patterned films of red-emitting CdSe/CdZnS ($\lambda_{em}$ = 622 nm), green-emitting CdSe/CdZnS ($\lambda_{em}$ = 526 nm), and blue-emitting CdS/ZnS ($\lambda_{em}$ = 460 nm) core/shell cQDs that were fabricated using template stripping. On the left hand side, the corresponding templates of the three films are shown. The absence of fluorescence from the central region of the template demonstrates that the films are stripped completely. The pattern transfer is highly uniform across large areas and shows high fidelity down to the submicron level, as is evidenced by both fluorescence microscopy imaging (Figure 1c) and high-resolution scanning electron microscopy (Figure 1d,e) of bull's-eye grating structures at the surface of a red-emitting cQD film.

The bull's-eye gratings shown in Figure 1c show a clear pitch-dependent intensity of the spontaneous emission. The substrate contains an array of gratings, each with 300 concentric circles and a 50% fill factor. These structures are patterned onto the surface of a red-emitting cQD film ($\lambda_{em}$ = 622 nm). From left to right, the grating pitch varies from $\Lambda$ = 550 to 300 nm with 50 nm decrements. The brightest emission in the low numerical aperture image (N.A. = 0.06) is observed for the circular grating with a pitch of $\Lambda$ = 400 nm, indicative of Bragg scattering of waveguided light along the surface normal. Before discussing the out-coupling enhancement and the improved brightness of our films, we first focus on the angular dependence of the Bragg scattering for the different pitches.

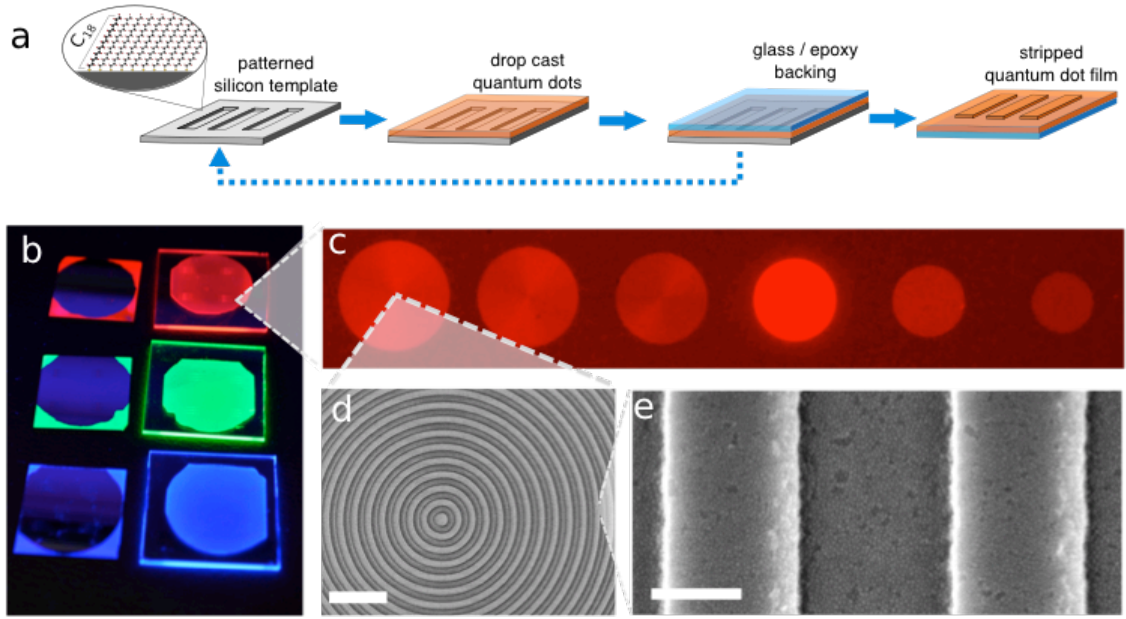

**Figure 1. (a)** Schematic of the template-stripping technique for fabricating surface-patterned cQD films. **(b)** Photograph of ultraviolet-illuminated patterned films of red-, green-, and blue-emitting cQDs that were fabricated using template stripping. On the left-hand side the corresponding templates of 2x2 cm each are shown. **(c)** Fluorescence microscopy image of an array of bull's-eye gratings of 300 concentric circles each, on a red-emitting cQD film. The pitch of these gratings varies from $\Lambda$ = 550 nm on the left, to $\Lambda$ = 300 nm on the right with 50 nm decrements. **(d, e)** Scanning electron micrographs at different magnifications of a bull's-eye grating with a pitch of 550 nm. Scale bars are 2 μm and 200 nm.

To obtain more insight into the angular dependence of the emission from the different bull's-eye gratings, we performed k-space imaging with high numerical aperture (N.A. = 0.8) light collection. Normalized k-space color maps taken at the center of the bull's-eye gratings of different pitches ($\Lambda$ between 500 and 300 nm) are shown in Figure 2a, and vertical line traces across the center of the k-space maps are shown in Figure 2b (for $\Lambda$ = 400, 450 and 500 nm). Indeed, the enhanced emission for the bull's-eye grating with pitch $\Lambda$ = 400 nm is centered at small k-vectors, indicating fluorescent beaming in the surface-normal direction. The resonant out-coupling occurs at different angles for both smaller and larger pitches, resulting in the observed donut-like emission patterns in k-space, with the pitch dependence of the angle of out-coupling following the Bragg condition given by equation (1) (see Figure S2). From the full-width-at-half-maximum (FWHM) of the $\Lambda$ = 400 nm line-trace in Figure 2b we measure the divergence of the beam to be 10.6°.

The width of the divergence is partially a result of spectral broadening of the resonance. The full spectral-angular response of the $\Lambda$ = 400 nm bull's-eye grating is shown in Figure 2c. This spectral-angular color-map is obtained by dispersing the k-space line trace of Figure 2b using an imaging spectrograph. It is clear that, despite the narrow emission linewidth of the cQD ensemble, distinct variations in the emission angle are present across the width of the emission spectrum. The shifting of the resonance as a function of wavelength can be fitted to equation (1), using the effective refractive index ($n_{eff}$) of the waveguided mode as the only fit parameter. Good agreement is found for $n_{eff}$ = 1.6 ± 0.1, as can be seen from the dashed white lines in Figure 2c. At the point of narrowest divergence, located at 638 nm, the divergence measures a FWHM of 6.1°. It is interesting to note that this value for spectrally-resolved divergence approaches those reported for cQD ensembles coupled to circular plasmonic gratings, where a divergence of 3.5 - 4.5° was recently reported.[34,35]

Aside from controlling the directionality of emission, our simple surface patterning strategy dramatically enhances the overall brightness of the cQD films. To quantify this, we compare the angular emission profile from the $\Lambda$ = 400 nm bull's-eye grating (see red line in Figure 2b) with the experimentally obtained angular emission profile from the non-resonant background. The out-coupling enhancement factor as a function of the collection cone is shown in Figure 2d. Integrated over a wide collection cone (> 100°), our patterned film emits almost twice as much as compared to the unpatterned case. Moreover, in the direct out-of-plane direction, the strong beaming effect results in over six-fold enhancement of out-coupling.

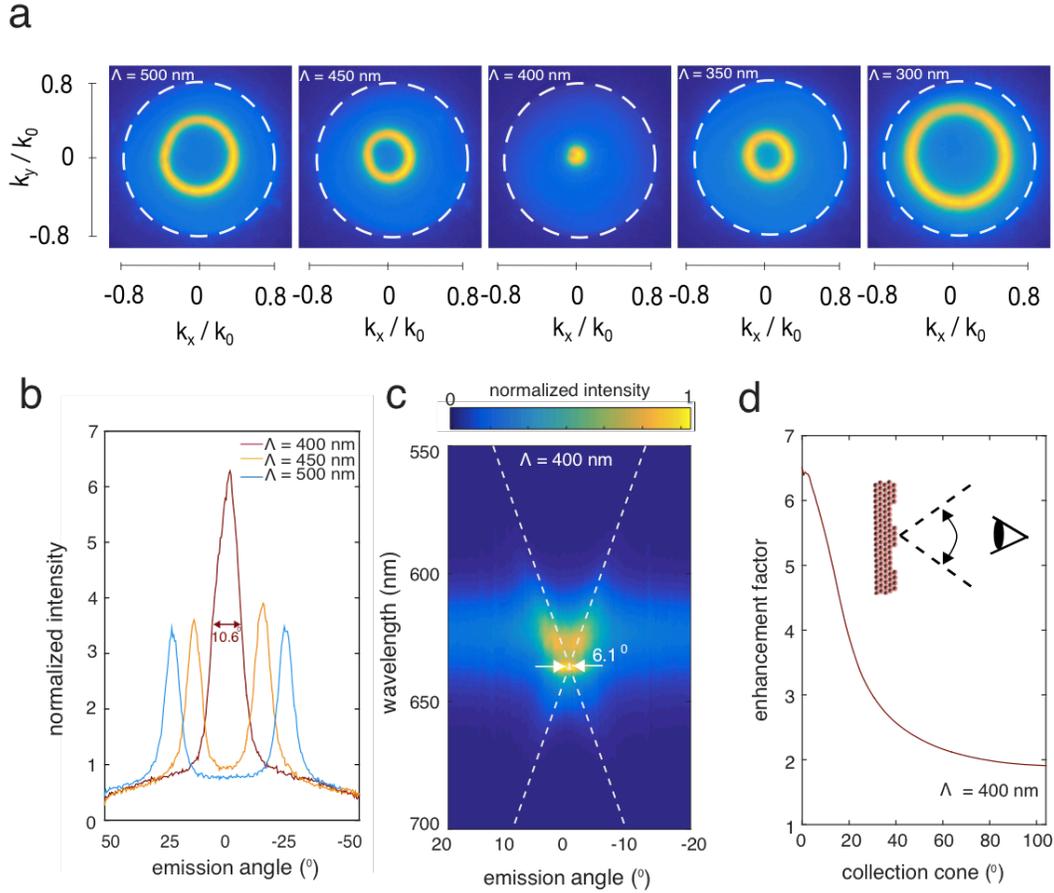

**Figure 2 (a)** Normalized k-space color maps of fluorescent emission from the right-most five bull's-eye gratings of Figure 1c ($\Lambda$ = 500 to 300 nm). Dashed white lines indicate the numerical aperture of the objective (N.A. = 0.8). **(b)** Vertical line traces at $k_x/k_0 = 0$ through the k-space color maps of the left most three panels of Figure 2a ($\Lambda$ = 500, 450, and 400 nm). Traces are normalized to the maximum non-resonant intensity. **(c)** Normalized spectral-angular color map of the $\Lambda$ = 400 nm bull's eye grating. Dashed white lines are fits to equation 1 (see text for details). **(d)** Enhancement of fluorescence out-coupling from the $\Lambda$ = 400 nm bull's eye grating as compared to the non-resonant case, as a function of the collection cone (see inset for schematic).

Beyond the out-coupling enhancement of spontaneous emission, our bull's-eye gratings also enable significantly improved stimulated emission characteristics. As previously mentioned, Bragg grating structures can provide both enhanced out-coupling through the first-order diffraction, as well as feedback in the in-plane direction through second-order diffraction. Indeed, under increasing excitation power we observe a clear transition from spontaneous emission to stimulated emission. This transition is characterized by a distinct spectral narrowing and sharp increase in the output power of the emission, an example of which is shown in Figure 3a for a red-emitting $\Lambda$ = 374 nm bull's-eye grating. The spectral narrowing results in a single-mode emission peak at 638 nm with a line-width of 0.7 nm (2 meV) and a total intensity increase of over an order of magnitude before reaching saturation (see inset of Fig. 3a). Spectral changes are accompanied by a reduction in the beam divergence from around 10° (174 mrad) for spontaneous emission down to 0.6° (10 mrad) for lasing, as can be seen from the k-space map in Figure 3b. The lasing from these structures exhibits low thresholds, consistently below 150 µJ/cm$^2$ and in some cases as low as 110 µJ/cm$^2$. These thresholds are comparable to those obtained using more complex microcavity structures or external distributed feed-back schemes.[29,36,37] Low-threshold lasing is indicative of efficient in-plane feedback and low-loss waveguiding, and is therefore a further confirmation of the high structural fidelity of our patterned films.

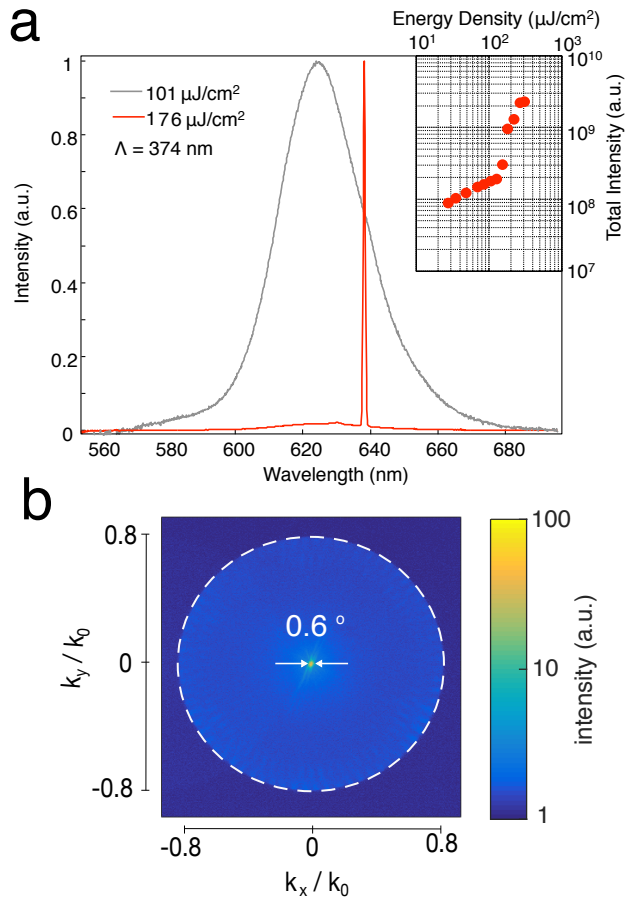

**Figure 3 (a)** Normalized emission spectra of a red-emitting $\Lambda$ = 374 nm bull's-eye grating below (grey line) and above (red line) the stimulated emission threshold. Inset: Total emission intensity as a function of energy density of the pump excitation, showing a clear threshold for stimulated emission at 120 μJ/cm$^2$ for this structure. **(b)** k-space color map on a logarithmic intensity scale of the emission of a bull's-eye grating above the lasing threshold, indicating the FWHM divergence of 0.6°. Dashed white lines indicate the numerical aperture of the objective (N.A. = 0.8).

A unique feature of the wavelength-scale patterning of cQD films is the combined spectral tunability of both the cQD building block as well as the photonic grating structure. As a demonstration of this flexibility, we fabricated bull's eye gratings with different pitches onto a mixed red-green-blue (RGB) cQD film to provide color tuning of spontaneous emission through spectrally selective out-coupling. A fluorescence image of the resulting structures is shown in Figure 4a. The emission from the unpatterned film in the background consists of a mix of the three individual colors to make up the spectrum displayed in Figure 4b. Each of the patterned Bragg gratings enhances a select portion of the emission spectrum of the mixed film in the surface-normal direction, leading to the varying colors observed in Figure 4a and the corresponding spectra shown in Figure 4b. Effectively, the different patterns shift the color gamut of the out-coupled light from the mixed RGB film (see Supporting Information for more details). The green and red color channels are most widely tunable, while tuning of blue channel is less effective. This is perhaps not surprising since the higher energy blue emission in the mixed film experiences more losses due to reabsorption by red and green dots. Reabsorption leads to reduced waveguiding, less interference, and thus less enhancement of the out-coupling through the first-order Bragg diffraction. A potential solution to reabsorption losses is to use Stoke-shift engineered cQDs,[38] where significant absorption for each color only occurs at wavelengths shorter than 450 nm (i.e. at energies above the blue emission spectrum). Nevertheless, as shown in Figure 4b, significant color shifts can be achieved for the different Bragg patterns with the current mixed films.

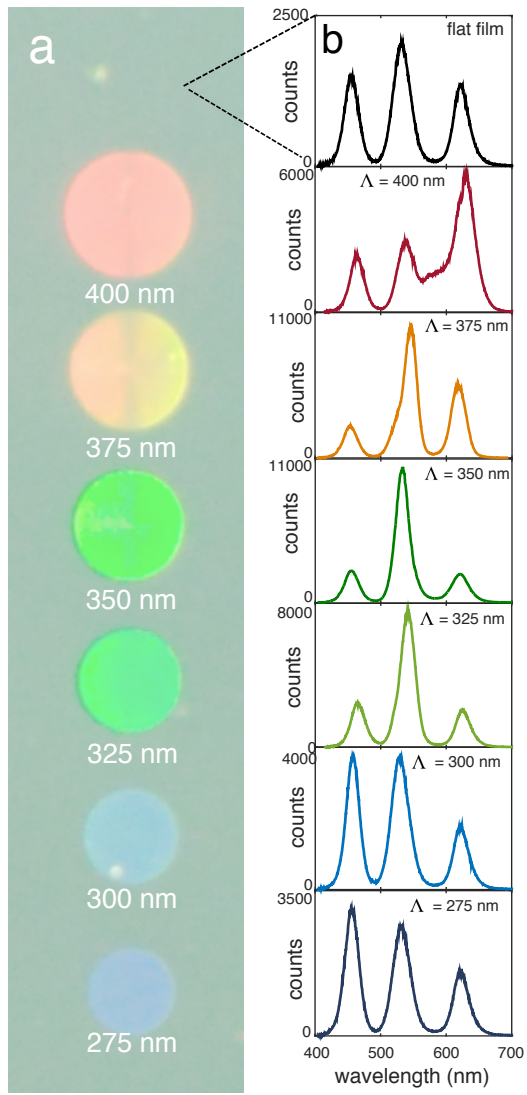

**Figure 4. (a)** Fluorescence microscopy image of an array of bull's-eye gratings patterned into a RGB-emitting mixed cQD film with 300 rings each and a pitch decreasing from top to bottom in 25 nm steps from $\Lambda$ = 400 to 275 nm. **(b)** Emission spectra of the unpatterned and patterned RGB-emitting mixed cQD film (N.A. = 0.2).

The demonstrated spatially and spectrally selective out-coupling of emission provides new levels of control over the optical properties of cQD assemblies. While previous strategies required complex photonic crystal[13] or plasmonic[39] structures to achieve such photonic control, the presented method of direct patterning provides a straightforward route towards enhanced and directional out-coupling of emission. In particular, the out-coupling enhancement of spontaneous and stimulated emission from patterned films has direct technological relevance in the use of cQD assemblies in optical down-conversion for backlight displays[11], LEDs[6], and single-mode surface-emitting lasers.[40]

**Methods.** *Template fabrication* – Patterned silicon (100) templates were fabricated using standard e-beam lithography and reactive ion etching techniques (see Supporting Information for details). Patterned templates were cleaned using $O_2$ plasma (600W, 3 min.) and piranha treatment (1:1 $H_2O_2/H_2SO_4$, 15 min.). The templates were subsequently coated with a dense self-assembled monolayer (SAM) using a previously reported octadecyltrichlorosilane treatment.[32] The resulting hydrophobic surface yields a contact angle > 110°.

*Template stripping* – Red-emitting CdSe/CdZnS ($\lambda_{em}$ = 622 nm), green-emitting CdSe/CdZnS ($\lambda_{em}$ = 526 nm), and blue-emitting CdS/ZnS ($\lambda_{em}$ = 460 nm) core shell quantum dots were synthesized based on previously reported procedures (see Supporting Information for details).[41–44] All three cQD systems have a comparable total core-shell particle diameter of approximately 10-12 nm. Special care was taken in the isolation of the quantum dots to remove unreacted material and excess ligands, since this significantly improves the homogeneity of drop casting in the following step and the ability to clean the templates for future use. cQDs were drop-cast from 9:1 hexane/octane solutions with optical densities (1 cm optical path length) at the first excitonic peak of 2.0 (red and blue) and 6.0 (green), using 35 μL for a 20 x 20 mm template size. Mixed RGB films were prepared using a ratio of 1:3:15 of red, green, and blue solutions in 9:1 hexane/octane of before mentioned optical densities. Template stripping was performed by placing a drop of epoxy (NOA 61, Norland Products) on the cQD film after drying, and gently placing a glass slide on top. The epoxy was allowed to flow towards the edges of the template until only the corners were free, after which it was cured with a 365 nm UV lamp for 30 min. After curing, the silicon template was separated from the cQD/epoxy/glass stack by placing a razor blade between the silicon template and the glass slide at one of the corners of the template, thereby exposing the patterned cQD film. Templates can be cleaned for future use through ultra-sonication in chloroform for 5 min. The SAM maintains sufficient quality for 10-15 drop casts, although best results are generally obtained with freshly prepared SAMs. An $O_2$ plasma (600W, 10 minutes) and piranha treatment (1:1 $H_2O_2/H_2SO_4$, 15 min.) can be used to remove the SAM, making the template ready for a new octadecyltrichlorosilane treatment.

*Fluorescence microscopy* – Optical characterization was performed using an inverted microscope (Nikon, TE200) equipped with a high-pressure mercury lamp (365 nm excitation). Fluorescence was collected using a 2x (N.A. 0.06) or 10x objective (N.A. 0.2), relay optics (focal length of f = 100 mm) and an imaging spectrograph (Horiba, Triax 320, 150 gr/mm grating) equipped with a nitrogen-cooled CCD detector (Princeton Instruments, Spec-10). Fluorescence microscopy images were recorded using a Nikon D3200 digital camera (white balance 3500K). For k-space imaging, fluorescence was collected using a 50x objective (Nikon, TU Plan Fluor, N.A. 0.8) and a 60 mm Fourier lens was used to project the back aperture of the objective onto the entrance slit of the imaging spectrograph. The k-space color maps were recorded using a fully opened slit and zero dispersion. Spectral-angular color maps were recorded using a 25 μm slit opening and spectral dispersion using a 150 gr/mm grating.

*Lasing experiments* – 450 nm pulses (~340 fs pulse duration, 1 KHz repetition rate) were generated by a collinear optical parametric amplifier (Spectra-Physics, Spirit-OPA) pumped by a 1040 nm pump laser (Spectra-Physics, Spirit-1040-8). After appropriate spectral filtering, the beam was directed through a gradient neutral density filter wheel to adjust the pulse power (Thorlabs, NDC-50C-2M-B). After beam expansion and collimation, a small portion of the beam was directed to a photodiode to monitor the pump power (Thorlabs, DET110). The rest of the beam was passed through a defocusing lens (f = 150 mm) into an inverted microscope (Nikon, Eclipse Ti-U). The beam was then directed upwards to the sample using a 488 nm long pass dichroic beamsplitter through a 50x air objective (Nikon, TU Plan Fluor, N.A. 0.8). The defocusing lens was adjusted to provide a spot size of ~90 um. The spot size was determined from the image of the photoluminescence from a flat portion of a film of template-stripped cQDs. A cross-section of the spot was then fitted with the sum of two Gaussian functions and the full-width-at-half-maximum was determined numerically. The excitation power density at the sample was monitored by correlating the power meter reading above the objective at the sample plane (Thorlabs, S170C with PM100D) with the current reading from the photodiode. Photoluminescence from the sample was collected by the same objective and directed through the dichroic beamsplitter, a 500 nm long pass emission filter, and relay lenses (f = 200 mm) into an imaging spectrometer (Andor, Shamrock 303i). For k-space imaging of the lasing, the lens for focusing into the spectrometer was replaced by a combination of an f = 50 mm lens and an f = 60 mm lens. The photoluminescence was dispersed with a 300 lines/mm grating (500 nm blaze) and imaged with an air-cooled electron-multiplying CCD camera (Andor, iXon 888 Ultra).

**ACKNOWLEDGEMENTS**

We gratefully acknowledge funding from the European Research Council under the European Union's Seventh Framework Program (FP/2007-2013)/ERC Grant Agreement 339905 (QuaDoPS Advanced Grant). F.P. acknowledges support by the Ambizione program of the Swiss National Science Foundation. J.C. acknowledges funding from the ETH Zurich Postdoctoral Fellowship Program and the Marie Curie Actions for People COFUND Program. We thank Y. Fedoryshyn, A. Olziersky, U. Drechsler, and M. Sousa for assistance in the fabrication and characterization.

**REFERENCES**

(1) Alivisatos, A. P. Semiconductor Clusters, Nanocrystals, and Quantum Dots. *Science* **1996**, *271*, 933–937

(2) Kim, J. Y.; Voznyy, O.; Zhitomirsky, D.; Sargent, E. H. 25th Anniversary Article: Colloidal Quantum Dot Materials and Devices: A Quarter-Century of Advances. *Adv. Mater.* **2013**, *25*, 4986–5010

(3) Chen, O.; Zhao, J.; Chauhan, V. P.; Cui, J.; Wong, C.; Harris, D. K.; Wei, H.; Han, H.-S.; Fukumura, D.; Jain, R. K.; et al. Compact high-quality CdSe-CdS core-shell nanocrystals with narrow emission linewidths and suppressed blinking. *Nat. Mater.* **2013**, *12*, 445–451

(4) Murray, C. B.; Norris, D. J.; Bawendi, M. G. Synthesis and characterization of nearly monodisperse CdE (E = sulfur, selenium, tellurium) semiconductor nanocrystallites. *J. Am. Chem. Soc.* **1993**, *115*, 8706–8715

(5) Shirasaki, Y.; Supran, G. J.; Bawendi, M. G.; Bulović, V. Emergence of colloidal quantum-dot light-emitting technologies. *Nat. Photonics* **2012**, *7*, 13–23

(6) Jang, E.; Jun, S.; Jang, H.; Lim, J.; Kim, B.; Kim, Y. White-light-emitting diodes with

quantum dot color converters for display backlights. *Adv. Mater.* **2010**, *22*, 3076–3080

(7) Kim, T.-H.; Cho, K.-S.; Lee, E. K.; Lee, S. J.; Chae, J.; Kim, J. W.; Kim, D. H.; Kwon, J.-Y.; Amaratunga, G.; Lee, S. Y.; et al. Full-colour quantum dot displays fabricated by transfer printing. *Nat. Photonics* **2011**, *5*, 176–182

(8) Colvin, V. L.; Schlamp, M. C.; Alivisatos, A. P. Light-emitting diodes made from cadmium selenide nanocrystals and a semiconducting polymer. *Nature* **1994**, *370*, 354–357

(9) Anikeeva, P. O.; Halpert, J. E.; Bawendi, M. G.; Bulović, V. Quantum Dot Light-Emitting Devices with Electroluminescence Tunable over the Entire Visible Spectrum. *Nano Lett.* **2009**, *9*, 2532–2536

(10) Shirasaki, Y.; Supran, G. J.; Bawendi, M. G.; Bulović, V. Emergence of colloidal quantum-dot light-emitting technologies. *Nat. Photonics* **2012**, *7*, 13–23

(11) Bourzac, K. Quantum dots go on display. *Nature* **2013**, *493*, 283

(12) Curto, A. G.; Volpe, G.; Taminiau, T. H.; Kreuzer, M. P.; Quidant, R.; van Hulst, N. F. Unidirectional emission of a quantum dot coupled to a nanoantenna. *Science* **2010**, *329*, 930–933

(13) Lodahl, P.; Floris Van Driel, A.; Nikolaev, I. S.; Irman, A.; Overgaag, K.; Vanmaekelbergh, D.; Vos, W. L. Controlling the dynamics of spontaneous emission from quantum dots by photonic crystals. *Nature* **2004**, *430*, 654–657

(14) Poitras, C. B.; Lipson, M.; Du, H.; Hahn, M. A.; Krauss, T. D. Photoluminescence enhancement of colloidal quantum dots embedded in a monolithic microcavity. *Appl. Phys. Lett.* **2003**, *82*, 4032

(15) Hoang, T. B.; Akselrod, G. M.; Argyropoulos, C.; Huang, J.; Smith, D. R.; Mikkelsen, M.

H. Ultrafast spontaneous emission source using plasmonic nanoantennas. *Nat. Commun.* **2015**, *6*, 7788

(16) Matterson, B. J.; Lupton, J. M.; Safonov, A. F.; Salt, M. G.; Barnes, W. L.; Samuel, I. D. W. Increased Efficiency and Controlled Light Output from a Microstructured Light-Emitting Diode. *Adv. Mater.* **2001**, *13*, 123–127

(17) Lupton, J. M.; Matterson, B. J.; Samuel, I. D. W.; Jory, M. J.; Barnes, W. L. Bragg scattering from periodically microstructured light emitting diodes. *Appl. Phys. Lett.* **2000**, *77*, 3340

(18) Erdogan, T.; Hall, D. G. Circularly symmetric distributed feedback semiconductor laser: An analysis. *J. Appl. Phys.* **1990**, *68*, 1435

(19) Jebali, A.; Mahrt, R. F.; Moll, N.; Erni, D.; Bauer, C.; Bona, G.-L.; Bächtold, W. Lasing in organic circular grating structures. *J. Appl. Phys.* **2004**, *96*, 3043

(20) Ziebarth, J. M.; Saafir, A. K.; Fan, S.; McGehee, M. D. Extracting Light from Polymer Light-Emitting Diodes Using Stamped Bragg Gratings. *Adv. Funct. Mater.* **2004**, *14*, 451–456

(21) Yang, J.; Choi, M. K.; Kim, D.-H.; Hyeon, T. Designed Assembly and Integration of Colloidal Nanocrystals for Device Applications. *Adv. Mater.* **2016**, *28*, 1176–1207

(22) Wood, V.; Panzer, M. J.; Chen, J.; Bradley, M. S.; Halpert, J. E.; Bawendi, M. G.; Bulović, V. Inkjet-Printed Quantum Dot-Polymer Composites for Full-Color AC-Driven Displays. *Adv. Mater.* **2009**, *21*, 2151–2155

(23) Kress, S. J. P.; Richner, P.; Jayanti, S. V; Galliker, P.; Kim, D. K.; Poulikakos, D.; Norris, D. J. Near-field light design with colloidal quantum dots for photonics and plasmonics. *Nano Lett.* **2014**, *14*, 5827–5833

(24) Kim, L.; Anikeeva, P. O.; Coe-Sullivan, S. A.; Steckel, J. S.; Bawendi, M. G.; Bulović, V. Contact printing of quantum dot light-emitting devices. *Nano Lett.* **2008**, *8*, 4513–4517

(25) Mentzel, T. S.; Wanger, D. D.; Ray, N.; Walker, B. J.; Strasfeld, D.; Bawendi, M. G.; Kastner, M. A. Nanopatterned electrically conductive films of semiconductor nanocrystals. *Nano Lett.* **2012**, *12*, 4404–4408

(26) Kim, J. Y.; Ingrosso, C.; Fakhfouri, V.; Striccoli, M.; Agostiano, A.; Curri, M. L.; Brugger, J. Inkjet-Printed Multicolor Arrays of Highly Luminescent Nanocrystal-Based Nanocomposites. *Small* **2009**, *5*, 1051–1057

(27) Tamborra, M.; Striccoli, M.; Curri, M. L.; Alducin, J. A.; Mecerreyes, D.; Pomposo, J. A.; Kehagias, N.; Reboud, V.; Sotomayor Torres, C. M.; Agostiano, A. Nanocrystal-Based Luminescent Composites for Nanoimprinting Lithography. *Small* **2007**, *3*, 822–828

(28) Sundar, V. C.; Eisler, H.-J.; Deng, T.; Chan, Y.; Thomas, E. L.; Bawendi, M. G. Soft-Lithographically Embossed, Multilayered Distributed-Feedback Nanocrystal Lasers. *Adv. Mater.* **2004**, *16*, 2137–2141

(29) Todescato, F.; Fortunati, I.; Gardin, S.; Garbin, E.; Collini, E.; Bozio, R.; Jasieniak, J. J.; Della Giustina, G.; Brusatin, G.; Toffanin, S.; et al. Soft-Lithographed Up-Converted Distributed Feedback Visible Lasers Based on CdSe-CdZnS-ZnS Quantum Dots. *Adv. Funct. Mater.* **2012**, *22*, 337–344

(30) Nagpal, P.; Lindquist, N. C.; Oh, S.-H.; Norris, D. J. Ultrasmooth patterned metals for plasmonics and metamaterials. *Science* **2009**, *325*, 594–597

(31) Alba, M.; Pazos-Perez, N.; Vaz, B.; Formentin, P.; Tebbe, M.; Correa-Duarte, M. A.; Granero, P.; Ferré-Borrull, J.; Alvarez, R.; Pallares, J.; et al. Macroscale Plasmonic Substrates for Highly Sensitive Surface-Enhanced Raman Scattering. *Angew. Chemie Int.*

*Ed.* **2013**, *52*, 6459–6463

(32) Lessel, M.; Bäumchen, O.; Klos, M.; Hähl, H.; Fetzer, R.; Paulus, M.; Seemann, R.; Jacobs, K. Self-assembled silane monolayers: an efficient step-by-step recipe for high-quality, low energy surfaces. *Surf. Interface Anal.* **2015**, *47*, 557–564

(33) McPeak, K. M.; van Engers, C. D.; Bianchi, S.; Rossinelli, A.; Poulikakos, L. V; Bernard, L.; Herrmann, S.; Kim, D. K.; Burger, S.; Blome, M.; et al. Ultraviolet Plasmonic Chirality from Colloidal Aluminum Nanoparticles Exhibiting Charge-Selective Protein Detection. *Adv. Mater.* **2015**, *27*, 6244–6250

(34) Harats, M. G.; Livneh, N.; Zaiats, G.; Yochelis, S.; Paltiel, Y.; Lifshitz, E.; Rapaport, R. Full spectral and angular characterization of highly directional emission from nanocrystal quantum dots positioned on circular plasmonic lenses. *Nano Lett.* **2014**, *14*, 5766–5771

(35) Livneh, N.; Harats, M. G.; Yochelis, S.; Paltiel, Y.; Rapaport, R. Efficient Collection of Light from Colloidal Quantum Dots with a Hybrid Metal–Dielectric Nanoantenna. *ACS Photonics* **2015**, *2*, 1669–1674

(36) Dang, C.; Lee, J.; Breen, C.; Steckel, J. S.; Coe-Sullivan, S.; Nurmikko, A. Red, green and blue lasing enabled by single-exciton gain in colloidal quantum dot films. *Nat. Nanotechnol.* **2012**, *7*, 335–339

(37) Roh, K.; Dang, C.; Lee, J.; Chen, S.; Steckel, J. S.; Coe-Sullivan, S.; Nurmikko, A. Surface-emitting red, green, and blue colloidal quantum dot distributed feedback lasers. *Opt. Express* **2014**, *22*, 18800–18806

(38) Coropceanu, I.; Bawendi, M. G. Core/shell quantum dot based luminescent solar concentrators with reduced reabsorption and enhanced efficiency. *Nano Lett.* **2014**, *14*, 4097–4101

(39) Livneh, N.; Strauss, A.; Schwarz, I.; Rosenberg, I.; Zimran, A.; Yochelis, S.; Chen, G.; Banin, U.; Paltiel, Y.; Rapaport, R. Highly directional emission and photon beaming from nanocrystal quantum dots embedded in metallic nanoslit arrays. *Nano Lett.* **2011**, *11*, 1630–1635

(40) Nurmikko, A. What future for quantum dot-based light emitters? *Nat. Nanotechnol.* **2015**, *10*, 1001–1004

(41) Reiss, P.; Bleuse, J.; Pron, A. Highly Luminescent CdSe/ZnSe Core/Shell Nanocrystals of Low Size Dispersion. *Nano Lett.* **2002**, *2*, 781–784

(42) Boldt, K.; Kirkwood, N.; Beane, G. A.; Mulvaney, P. Synthesis of Highly Luminescent and Photo-Stable, Graded Shell CdSe/Cd$_x$Zn$_{1-x}$S Nanoparticles by In Situ Alloying. *Chem. Mater.* **2013**, *25*, 4731–4738

(43) Bae, W. K.; Char, K.; Hur, H.; Lee, S. Single-Step Synthesis of Quantum Dots with Chemical Composition Gradients. *Chem. Mater.* **2008**, *20*, 531–539

(44) Lee, K.-H.; Lee, J.-H.; Song, W.-S.; Ko, H.; Lee, C.; Lee, J.-H.; Yang, H. Highly efficient, color-pure, color-stable blue quantum dot light-emitting devices. *ACS Nano* **2013**, *7*, 7295–7302

*Supporting Information for:*

# Direct Patterning of Colloidal Quantum-Dot Thin Films for Enhanced and Spectrally Selective Out-Coupling of Emission

Ferry Prins[*], David K. Kim[*], Jian Cui, Eva De Leo, Kevin M. McPeak, and

David J. Norris

Optical Materials Engineering Laboratory, Department of Mechanical and Process Engineering,

ETH Zurich, 8092 Zurich, Switzerland

* these authors contributed equally

To whom correspondence should be addressed: dnorris@ethz.ch (D.J.N.)

# 1. Material Synthesis

**Chemicals:** Cadmium oxide (CdO, 99.999%) was purchased from Strem Chemicals. n-dodecylphosphonic acid (DDPA, 98%) was purchased from Epsilon Chimie. 1-butanol (ACS Grade) was purchased from Merck KGaA. Diphenylphosphine (DPP, 98%), ethanol (96%), hexadecylamine (HDA, 90%), hexane (95%), methanol (99.9%), selenium pellets (99.999%), sulfur (99.5%), trioctylphosphine (TOP, 90% and 97%), trioctylphosphine oxide (TOPO, 90%), 1-octadecene (ODE, 90%), 1-octanethiol (98.5%), octylamine (99%), oleylamine (OAm, 70%), oleic acid (OLA, 90%), and zinc acetate [$Zn(ac)_2$, 99.99%] were purchased from Sigma.

Cadmium oleate stock and zinc oleate stock were prepared as previously reported[1] and stored in the glovebox for future use.

**Preparation of Red-Emitting Core/Shell Quantum Dots:** CdSe cores with a lowest energy exciton peak of 579 nm (size approximately 3.8 nm, as determined by Yu *et al.*[2]) were synthesized following a modified version of a procedure by Reiss *et al.*[3] Briefly, 822 mg of CdO (6.4 mmol), 16.2 g of TOPO (42 mmol), 37 g of HDA (153.2 mmol), and 3.215 g of DDPA (12.8 mmol) were combined in a 250 mL 4-neck round-bottom flask and heated to 90 °C under $N_2(g)$. Then, the mixture was degassed 3 times to below 0.1 Torr with stirring at 1000 rpm. The flask was filled with $N_2(g)$ and heated at 320 °C until the solution turned clear. The temperature was then reduced to 260 °C and 8 mL of 1M Se in TOP (97%) solution containing 85 μL of DPP was swiftly injected at 300 rpm. After injection, the stir rate was increased to 1000 rpm and the temperature was maintained at 260 °C. The growth of CdSe nanocrystals was monitored using an UV-visible spectrophotometer (Varian Cary 300, UV-VIS). The reaction was quenched by removing the heating mantle, cooling to 200 °C with an air gun, submerging the flask in a water bath, and adding 40 mL of 1-butanol at 130 °C to prevent solidification. To isolate the product, the reaction mixture was split into 6 centrifuge tubes and methanol was added to reach a total volume of 50 mL in each tube. A precipitate was collected by centrifugation at 4000 rpm for 10 min, redispersed in 20 mL of hexane, and left undisturbed overnight at room temperature (RT). The following day, the mixture was centrifuged at 4000 rpm for 20 min, the supernatant was transferred to 6 centrifuge tubes, and ethanol was added to precipitate the product. The nanocrystals were isolated by centrifugation at 4000 rpm for 10 min. The precipitate in each tube was dispersed in 4 mL of hexane to be precipitated out once more with ethanol and centrifuged under the same conditions. The nanocrystals were dispersed in a 20 mL of hexane and stored in the dark until further use.

The growth of the CdS/ZnS shell on CdSe nanocrystals was performed following Boldt *et al.*[4] and as previously reported by our group.[1] Briefly, 100 nmol of CdSe nanocrystals in hexane, 3 mL of ODE, and 3 mL of OAm were combined in a 3-neck 100 mL round-bottom flask and degassed at RT for 1 h. The mixture was then heated at 120 °C for 20 min. The flask was filled with $N_2$(g) and heated to 305 °C at a rate of 16 °C/min. At 210 °C, two separate syringes containing cadmium oleate (0.22 mmol) and octanethiol (0.26 mmol), each diluted in ODE (final volume of 3 mL), were injected with a syringe pump at a rate of 1.5 mL/h. After injection, the temperature was lowered to 200 °C, 1 mL of degassed OLA was added dropwise, and the mixture was allowed to anneal for 1 h at 200 °C. Subsequently, the temperature was lowered to 120 °C and the reaction mixture was degassed for 30 min. The flask was then filled with $N_2$(g) and the temperature was raised to 280 °C at a rate of 16 °C/min. At 210 °C, two separate syringes containing zinc oleate (0.24 mmol) and octanethiol (0.48 mmol) each diluted in ODE (final volume of 3 mL) were injected with a syringe pump at a rate of 2.5 mL/h. After injection, the reaction mixture was quickly cooled down to RT. To precipitate and clean the nanocrystals, an equivalent amount of ethanol to reaction-mixture volume was added and centrifuged at 4000 rpm for 10 min. The bright precipitate was dispersed in 4 mL of hexane, 10 mL of ethanol was added, and the mixture was centrifuged under the same conditions. This step was repeated (2 mL of hexane and 5 mL of ethanol). The product was dispersed in a small volume of hexane and stored in the dark until future use. The recipe resulted in core/shell nanocrystals with a lowest energy excitonic absorption peak at 607 nm and an emission peak at 622 nm with a full-width-at-half-maximum (fwhm) of 28 nm.

**Preparation of Green-Emitting Core/Shell Quantum Dots:** The core/shell CdSe/ZnS nanocrystals with a composition gradient were prepared by modifying a procedure by Bae *et al.*[5] Briefly, 51.4 mg of CdO (0.4 mmol), 734 mg of Zn(ac)$_2$ (3.34 mmol), and 2 mL of OLA (6 mmol) were combined in a 3-neck 100 mL round-bottom flask and degassed 3 times under 0.1 Torr. The mixture was heated at 150 °C and degassed for 30 min. Then, the flask was filled with N$_2$(g), 15 mL of degassed ODE was added and degassed at 120 °C for an additional 5 min. Subsequently, the mixture was heated at 310 °C to form a clear solution. 3 mL of TOP (90%) containing 0.2 mmol Se and 4 mmol S was then swiftly injected into the reaction mixture. The temperature was maintained at 300 °C for 10 min and it was then quickly cooled down to RT. The reaction mixture was transferred to a centrifuge tube, filled with ethanol, vortexed and centrifuged at 4000 rpm for 10 min. This ethanol extraction procedure was repeated (typically 5 times) until a bright pellet was fully precipitated with a clear supernatant. The product was dispersed in 8 mL of hexane and stored in the dark until future use. The recipe resulted in core-shell nanocrystals with a lowest energy excitonic absorption peak at 530 nm and an emission peak at 545 nm.

To grow a ZnS shell on CdSe/ZnS nanocrystals a modified recipe by Boldt *et al.* was followed. Briefly, 100 nmol of CdSe/ZnS nanocrystals in hexane (which was approximated with an absorption cross-section of 517 nm CdSe cores), 3 mL of ODE, 3 mL of OAm, and 1 mL OLA were combined in a 3-neck 100 mL round-bottom flask and degassed at room temperature for 1 h. The mixture was heated to 120 °C for 20 min, switched to N$_2$(g) and raised to 280 °C at a rate of 16 °C/min. When 210 °C was reached, two separate syringes containing zinc oleate (0.25 mmol) and octanethiol (0.5 mmol), each diluted in ODE (final volume of 3 mL), were injected using a syringe pump at a rate of 2.5 mL/h. After injection, the temperature was lowered to 200 °C and 1 mL of degassed OLA was added dropwise. The reaction annealed at 200 °C for 1 h, and then it was quickly cooled down to RT. The nanocrystals were purified and isolated by precipitation with ethanol and centrifugation at 4000 rpm for 10 min. The resulting bright precipitate was dispersed in 2 mL of hexane 5 mL of ethanol was slowly added, and the mixture was centrifuged once more. This step was repeated (1 mL hexane and 3 mL ethanol). The product was dispersed in 3 mL of hexane and stored in the dark until future use. The recipe resulted in nanocrystals with a lowest energy excitonic absorption peak at 505 nm and an emission peak at 526 nm with a fwhm of 29 nm.

**Preparation of Blue-Emitting Core/Shell Quantum Dots:** The core/shell nanocrystals with a composition gradient were prepared by modifying a procedure by Lee *et al.*[6] Briefly, 128.4 mg of CdO (1 mmol), 1832 mg of Zn(ac)$_2$ (8.3 mmol), and 7 mL of OLA (22 mmol) were combined in a 3-neck 100 mL round-bottom flask. The mixture was degassed 3 times under 0.1 Torr and at 150 °C for 20 min while stirring at 1000 rpm. The flask was filled with $N_2(g)$, 15 mL of degassed ODE was added, and the mixture was degassed at 120 °C for 10 min. The flask was filled with $N_2(g)$ and heated to 305 °C to obtain a clear solution. At 305 °C, a clear solution consisting of 3 mL of ODE with 1.6 mmol of S prepared inside the glovebox, was swiftly injected into the reaction mixture. After injection, the temperature was maintained at 305 °C for 12 min. Subsequently, a clear solution consisting of 5 mL of OLA with 4 mmol of sulfur prepared in the glovebox, was injected with a syringe pump at a rate of 0.5 mL/min. After injection, the temperature was maintained at 305 °C for 3 h. After annealing, the reaction mixture was quickly cooled down to RT. To precipitate and purify the nanocrystals, the reaction mixture was split into two centrifuge tubes. Then 20 mL of ethanol was added to each tube and centrifuged at 8000 rpm for 10 min. The solid was dispersed in 5 mL of hexane, 1 mL of octylamine was added, and the mixture was sonicated for 5 min. Then, 20 mL of ethanol was added and the mixture was centrifuged at 8000 rpm for 10 min. The cleaning process was repeated 2 times. The particles were cleaned 3 times using a combination of hexane (4 mL) and a mixture of acetone (10 mL) ethanol (20 mL), with a clear supernatant being discarded each time. The particles were dispersed in 5 mL of hexane and stored in the dark until future use. The recipe resulted in core/shell nanocrystals with a lowest energy excitonic absorption peak at 450 nm and an emission peak at 460 nm with a fwhm of 25 nm.

## 2. Sample Fabrication

**Materials:** Four-inch diameter, single-side-polished, single-crystalline Si(100) wafers with <0.4 nm root-mean-square (RMS) roughness and thicknesses of either 500 or 1000 μm were purchased from Silicon Valley Microelectronics and diced into 2 x 2 cm square pieces. Octadecyltrichlorosilane (OTS, 96%) was purchased from Merck KGaA. Bicyclohexyl (99%, D79403), carbon tetrachloride ($CCl_4$, 99.5%, 289116), chloroform (99.8%, 132950), and sulfuric acid ($H_2SO_4$, 95-98%, 258105) were purchased from Sigma. Hydrogen peroxide ($H_2O_2$, 30%, AnalaR Normapur) was purchased from VWR. Methyl isobutyl ketone (MIBK, Technic France Micropur, VLSI Grade) and isopropanol (IPA, Technic France Micropur, VLSI Grade) was provided by IBM Zurich's Binnig and Rohrer Nanotechnology Center (BRNC).

**Template Fabrication:** Si(100) substrates of 2 x 2 cm in size were cleaned using sonication in acetone, isopropanol, and deionized (DI) water and prebaked at 180 °C for 10 min. Substrates were coated with 200 nm PMMA (AR-P 672.03 950K) and patterns were defined using electron-beam lithography (Vistec, NFL 5). Patterns were developed using MIBK:IPA (1:2 ratio) for 2 min, followed by 1 min in fresh IPA and then 1 min under running DI water. Features 100 nm deep were etched into the Si substrate using deep-reactive-ion etching (Alcatel, AMS-200 I Speeder) at 11.25 mTorr with 60 SCCM $C_4F_8$, 40 SCCM $SF_6$, a plate power of 80 W, and an inductive-coupled-plasma (ICP) power of 1200 W. The etch rate for these settings was ~8 nm/s. The PMMA mask was removed using ultrasonication in acetone, followed by an IPA rinse, and subsequent $O_2$ plasma at 750 mTorr, 600 SCCM $O_2$, and 600 W (PVA-TePla, GIGAbatch 310 M) for 3 min.

**Surface Functionalization:** Patterned Si substrates were functionalized with octadecyltrichlorosilane (OTS) based self-assembled monolayers following a recipe by Lessel *et al.*[7] In short, silicon templates were cleaned in a piranha solution (1:1 $H_2O_2$, $H_2SO_4$) for 15 min, submerged in DI water for 5 min, then blow dried with a $N_2$ gun. In a Teflon container, 150 μL of OTS, 300 μL of $CCl_4$, and 10 mL of bicyclohexyl were combined. A beaker of DI water was set to 170 °C on a hotplate, and the silicon substrate was quickly passed over the water vapor. Immediately after the thin film of water evaporated, the sample was placed in the Teflon container with OTS for 15 min. Next, to remove non-bound OTS, the samples were thoroughly rinsed with cholorform and subsequently sonicated in a chloroform bath for 2 min. Samples were placed back in the Teflon container with OTS, rinsed and sonicated two more times to complete the process. The resulting hydrophobic surface yields a contact angle of > 110° for unpatterned areas.

**Quantum-Dot Deposition:** The concentration was determined by taking the optical density of the quantum-dot dispersion with a UV-VIS spectrophotometer at the lowest energy exciton (red = 607 nm, green = 505 nm, blue = 450 nm). For the red-emitting quantum dots, we prepared the cQD dispersion at an optical density of 2.0 (1 cm optical path length). For the green-emitting quantum dots, we prepared the cQD dispersion at an optical density of 6.0. For the blue-emitting quantum dots, we prepared the cQD dispersion at an optical density of 2. All dispersions were prepared using a hexane:octane ratio of 9:1. Then, 35 µL of cQDs was dropcast onto the silicon substrate and allowed to dry, yielding a uniform film approximately 200 nm in thickness. For white light, we mixed 10 µL of red-emitting quantum dots, 30 µL of green-emitting quantum dots, 150 µL of blue-emitting quantum dots, using the before mentioned optical densities with 9:1 hexane:octane ratio. The apparent overexpression of green and blue accommodates for energy-transfer dynamics and the resulting red-shift of spectral features in the film. 35 µL of red-green-blue-emitting (RGB) cQDs was dropcast onto the silicon substrate and dried to yield a uniform film. Template stripping was then performed by placing a drop of epoxy (NOA 61, Norland Products) on the cQD film after drying. The epoxy was allowed to flow towards the edges of the template until only the corners were free, after which it was cured with a 365 nm UV lamp for 30 min. Please note that, although minor percolation of the epoxy into possible void spaces of the cQD film cannot be excluded, no effect on the quality and effective refractive index of the film was observed. After curing, the silicon template was separated from the QD/epoxy/glass stack by placing a razor blade between the silicon template and the glass slide at one of the corners of the template, thereby exposing the patterned cQD film. Templates can be cleaned for future use through ultra-sonication in chloroform for 5 min. The SAM maintains sufficient quality for 10-15 dropcasts, although best results are generally obtained with freshly prepared SAMs. An $O_2$ plasma (600W, 10 min) and piranha treatment (1:1 $H_2O_2/H_2SO_4$, 15 min) can be used to remove the SAM, making the template ready for a new octadecyltrichlorosilane treatment.

## 3. Atomic Force Microscopy (AFM)

Atomic force microscopy (Bruker FastScan) was performed to characterize the height of the cQD grating structure. Figure S1 shows an atomic force micrograph of a linear grating on a red-emitting quantum-dot film with a pitch of 550 nm. The height of the pattern is consistent with the etch depth of the template, which was set to 100 nm.

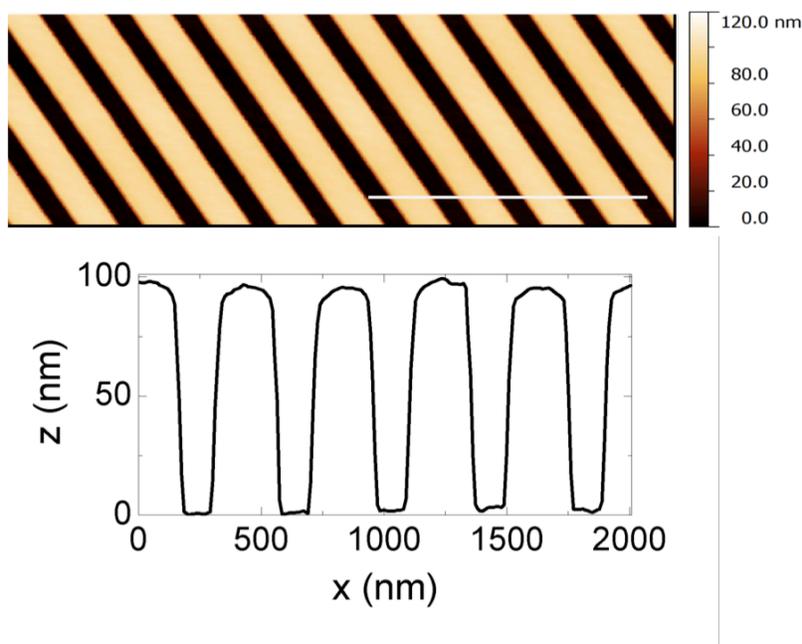

**Figure S1.** Top: Atomic force micrograph of a cQD linear grating with 400 nm pitch. Bottom: line trace taken perpendicular to the same grating structure.

## 4. Angular Dependence of Emission for Varying $\lambda_{em}$

Figure S2 shows the angle of emission from different circular Bragg gratings for red-, green-, and blue-emitting cQD films as a function of pitch. The measured emission angles can be fit to Equation (1) in the main text, using the effective refractive index of the waveguided mode as the only fitting parameter. We obtain $n_{eff}$ values of 1.58 ± 0.02, 1.55 ± 0.04 and 1.73 ± 0.02 for the red-, green-, and blue-emitting films, respectively. Note that the $n_{eff}$ value for the red-emitting film that is obtained here is consistent with the one obtained from the fit in Figure 2c in the main text. The effective refractive index of our waveguided modes is a non-trivial quantity that depends sensitively on the geometry of the structured surface, as well as the composition and fill-factor of the cQD film. Generally, the increasing effective refractive index values in the blue wavelength region are consistent with previous reports in literature.[8]

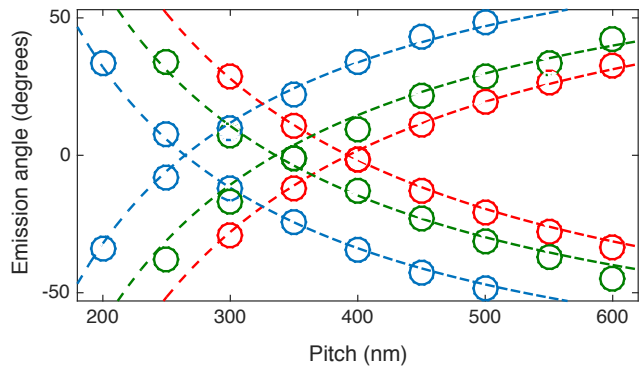

**Figure S2**. Emission angle with respect to the surface normal for different circular Bragg gratings for red, green, and blue cQD films as a function of the pitch of the grating. Open circles indicate experimental data for three wavelengths with associated uncertainty that is smaller than the symbol size. The dashed lines are fits to equation 1 of the main text.

## 5. Color Gamut Tuning Through Surface Patterning of Mixed Films

We fabricated bull's-eye gratings with different pitches onto a mixed RGB cQD film to provide color tuning of spontaneous emission through spectrally selective out-coupling (see Figure 4 in the main text). The different patterns shift the color gamut of the out-coupled light from the mixed RGB film. To quantify the tunability of color through the surface patterning, we plot the color coordinates of each spectrum on a International Commission on Illumination (CIE) color map, as shown in Figure S3.

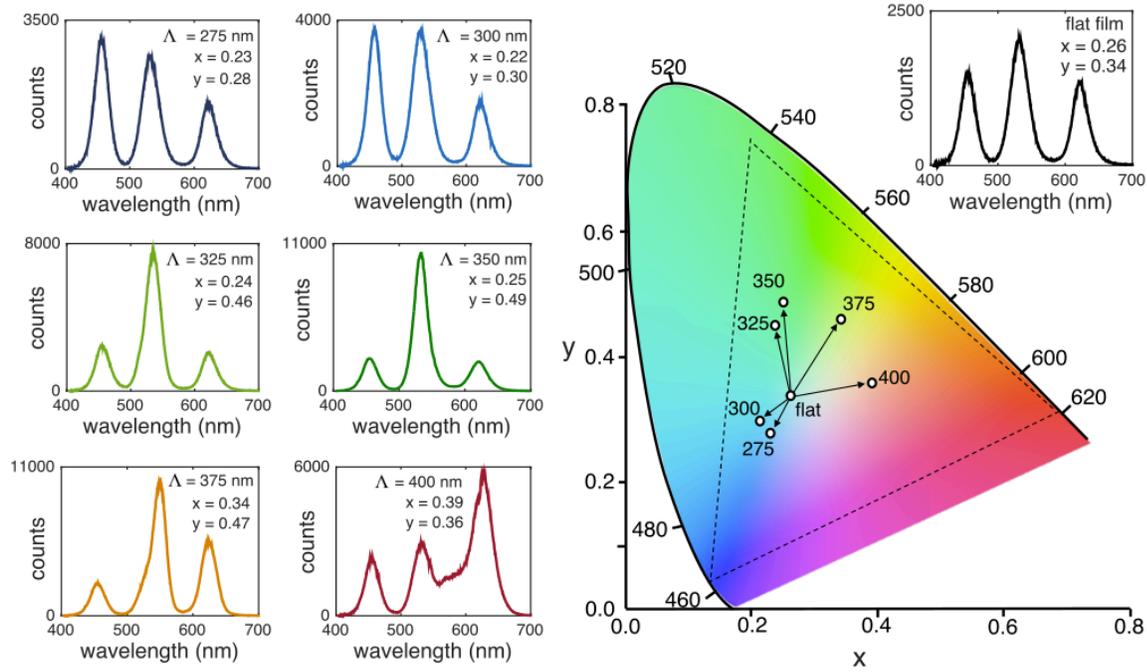

**Figure S3**. Emission spectra of the unpatterned and patterned RGB-emitting mixed-cQD film recorded using a N.A. = 0.2 objective indicating the CIE x and y values, and CIE color map indicating the tunability of the color gamut of the out-coupled emission.

**Supplementary References**

(1) Kress, S. J. P.; Antolinez, F. V; Richner, P.; Jayanti, S. V; Kim, D. K.; Prins, F.; Riedinger, A.; Fischer, M. P. C.; Meyer, S.; McPeak, K. M.; et al. Wedge Waveguides and Resonators for Quantum Plasmonics. *Nano Lett.* **2015**, *15*, 6267.

(2) Yu, W. W.; Qu, L.; Guo, W.; Peng, X. Experimental Determination of the Extinction Coefficient of CdTe, CdSe, and CdS Nanocrystals. *Chem. Mater.* **2003**, *15*, 2854.

(3) Reiss, P.; Bleuse, J.; Pron, A. Highly Luminescent CdSe/ZnSe Core/Shell Nanocrystals of Low Size Dispersion. *Nano Lett.* **2002**, *2*, 781.

(4) Boldt, K.; Kirkwood, N.; Beane, G. A.; Mulvaney, P. Synthesis of Highly Luminescent and Photo-Stable, Graded Shell CdSe/Cd$_X$Zn$_{1-X}$S Nanoparticles by In Situ Alloying. *Chem. Mater.* **2013**, *25*, 4731.

(5) Bae, W. K.; Char, K.; Hur, H.; Lee, S. Single-Step Synthesis of Quantum Dots with Chemical Composition Gradients. *Chem. Mater.* **2008**, *20*, 531.

(6) Lee, K.-H.; Lee, J.-H.; Song, W.-S.; Ko, H.; Lee, C.; Lee, J.-H.; Yang, H. Highly Efficient, Color-Pure, Color-Stable Blue Quantum Dot Light-Emitting Devices. *ACS Nano* **2013**, *7*, 7295.

(7) Lessel, M.; Bäumchen, O.; Klos, M.; Hähl, H.; Fetzer, R.; Paulus, M.; Seemann, R.; Jacobs, K. Self-Assembled Silane Monolayers: An Efficient Step-by-Step Recipe for High-Quality, Low Energy Surfaces. *Surf. Interface Anal.* **2015**, *47*, 557.

(8) Diroll, B. T.; Gaulding, E. A.; Kagan, C. R.; Murray, C. B. Spectrally-Resolved Dielectric Functions of Solution-Cast Quantum Dot Thin Films. *Chem. Mater.* **2015**, *27*, 6463.